\documentclass{appolb}
\usepackage{amssymb}
\usepackage{epsfig}
\usepackage{amsmath}
\usepackage{graphicx}
\usepackage{epsfig}
\usepackage{amsmath}
\usepackage{amssymb}
\newcommand{\w}{\vrule height 13 pt depth 0 pt width 0 pt}

\begin{document}
\title{SU(3) Breaking in Decays of Exotic Baryons}
\author{Micha{\l} Prasza{\l}owicz
\address{Nuclear Theory Group,
Brookhaven National Laboratory\\
Upton, NY 19973-5000, USA\\
e-mail: {\tt michal@quark.phy.bnl.gov}\\
and\\
 M. Smoluchowski Institute of Physics,
 Jagellonian University\\ 
Reymonta 4, 30-059 Kraków, Poland}}
\maketitle

\begin{abstract}
Within the chiral soliton model the SU(3) breaking
collective Hamiltonian introduces representation mixing in the
baryonic wave functions. We calculate ${\cal{O}}(m_s)$ effects of
this mixing on the decay widths of decuplet and antidecuplet
baryons. We find importance of the 27-plet admixture in the
${\it\Theta}^+$ and ${\it\Xi}_{\overline{10}}$ decays. The role of the
$1/N_{\rm c}$ nonleading
 terms in ${\cal{O}}(m_s)$ transition matrix elements is
 discussed.
\end{abstract}
\section{Introduction}

\label{intro}
\vspace{0.3cm}
There is almost no doubt today that the lightest member of the exotic
antidecuplet has been discovered \cite{exp}. Most probably also the heaviest
members of $\overline{10}$ were seen by NA49 experiment at CERN \cite{Xi}.
These states were predicted within the chiral soliton models
\cite{antidec,Mogil,DPP,Weigel}. Early estimates of \emph{both} ${\it\Theta}
^{+}$ and ${\it\Xi} _{\overline{10}}$ masses from the second order mass formulae
obtained in the Skyrme model are in a surprising agreement with present
experimental findings \cite{Mogil}. Later, the masses, as well as the decay
widths of the exotic states were computed within the chiral quark soliton
model \cite{DPP}. There, however, a freedom in relating the exotic-nonexotic
splittings and the splittings within the exotic multiplets to the value of
the pion--nucleon sigma term, ${\it\Sigma} _{\pi N}$, whose experimental value has
varied over the years from 45 to almost 77 MeV \cite{sigma}, made the
prediction of ${\it\Xi} _{\overline{10}}$ higher \cite{DPP} than the 1860 MeV
reported in \cite{Xi}. That the chiral models can easily accommodate lighter
${\it\Xi} _{\overline{10}}$ masses is clear from earlier studies \cite%
{Mogil,other27,Ma} and was emphasized recently in Ref.~\cite{EKP}.

\newpage
One of the most striking predictions of the seminal paper by Diakonov,
Petrov and Polyakov \cite{DPP} was the narrow width of antidecuplet states.
Despite some misprints in this paper (see \eg \cite{EKP,Arndt}) and the
model dependent corrections, the narrow width is one of the key features of
the chiral model predictions which is in line with recent experimental
findings. The small width appears due to the cancellation of the coupling
constants multiplying $3$ different group-theoretical structures entering
the decay operator \cite{DPP}. That this cancellation is consistent with the
$N_{\rm c}$ counting, despite the fact that two out of the three above mentioned
constants scale differently with $N_{\rm c}$ was shown in Ref.~\cite{MPGamma}.

In Ref.~\cite{DPP} also the $m_{s}$ corrections to the decay widths coming
from the representation mixing in the baryonic wave functions, caused by the
SU(3) breaking effective Hamiltonian, were estimated. However, two
approximations have been used: firstly only the mixing with lowest possible
representations was considered, boiling down to neglecting the SU(3)-flavor $%
27$-plet; and secondly, these corrections were calculated only for the term
leading in $N_{\rm c}$. In Ref.~\cite{Arndt} the second simplification was
partially abandoned and in Ref.~\cite{Ma} the $27$-plet contributions were
evaluated, however, only for the leading term. It is the purpose of this
work to discuss the $m_{s}$ corrections to the decay widths without the
above mentioned simplifications. Some of the results presented here were
already discussed in Ref.~\cite{EKP}.

The corrections due to the representation mixing constitute only a part of
the full $m_{s}$ correction which, however, are fully under control if the
mass spectra are known. There exists another set of $m_{s}$ corrections in
the decay operator itself. The group theoretical structure of these terms is
known \cite{oper}, however, numerical analysis is not straightforward, since
there are 3 new unknown constants which appear at this order. In the
following we concentrate only on the mixing corrections. Therefore their
magnitude can only serve as an estimate of a theoretical uncertainty
introduced by $m_{s}$ corrections.

Our findings can be summarized as follows: $27$-plet admixtures are very
important for the ${\it\Theta}^{+}$ and ${\it\Xi}_{\overline{10}}$ decays, whereas for
${\it\Sigma}_{\overline{10}}$ decays they are only moderate and for $N_{%
\overline {10}}$ decays the can be safely neglected. Effects of the terms
nonleading in $N_{\rm c}$ are important for ${\it\Theta}^{+}$ decay: in fact they
change the character of the correction from enhancement found in \cite{DPP}
to suppression discussed already in \cite{EKP}.

The paper is organized as follows: in Sect.~\ref{mixing} we introduce model
parameters and discuss the magnitude of the representation mixing. In Sect.~%
\ref{decays} we calculate decay corrections to the decuplet and antidecuplet
decay widths. We explicitly display their dependence on the model parameter $%
\rho$ and pion-nucleon sigma term, and examine the importance of the $27$%
-plet. Conclusions are presented in Sect.~\ref{summ}.

\section{Representation mixing}

\label{mixing}

In a recent paper \cite{EKP} it has been shown that the set of parameters of the
symmetry breaking Hamiltonian%
\begin{equation}
\hat{H}^{\prime}=\alpha D_{88}^{(8)}+\beta Y+\frac{\gamma}{\sqrt{3}}%
D_{8i}^{(8)}\hat{S}_{i}  \label{Hprim}
\end{equation}

\noindent
(where $D_{88}^{(8)}$ are SU(3) Wigner matrices, $Y$ is hypercharge and $%
\hat{S}_{i}$ is the collective spin operator \cite{Blotzsu3}) which
reproduces well the nonexotic spectra, as well as the measured mass of the $%
{\it\Theta}^{+}(1540)$ can be parameterized as follows\footnote{%
We use here $m_{s}/(m_{u}+m_{d})=12.9$ \cite{Leutwyler}.}:%
\begin{equation}
\alpha=336.4-12.9\,{\it\Sigma}_{\pi N}\,,\quad\beta=-336.4+4.3\,{\it\Sigma}_{\pi
N},\quad\gamma=-475.94+8.6\,{\it\Sigma}_{\pi N}\,.  \label{albega}
\end{equation}
Moreover, the inertia parameters which describe the representation splittings%
\begin{equation}
{\it\Delta} M_{{10}-8}=\frac{3}{2I_{1}},\;\;{\it\Delta} M_{{\overline{10}}-8}=\frac{3}{%
2I_{2}}
\end{equation}
take the following values (in MeV)%
\begin{equation}
\frac{1}{I_{2}}=152.4\,,\quad\frac{1}{I_{2}}=608.7-2.9\,{\it\Sigma}_{\pi N}\,.
\end{equation}
If, furthermore, one imposes additional constraint that $M_{{\it\Xi} _{\overline{%
10}}}=1860$ MeV, then ${\it\Sigma}_{\pi N}=73$ MeV \cite{EKP} (see also \cite%
{Schw}) in agreement with recent experimental estimates \cite{sigma}. The
dependence of model parameters on ${\it\Sigma}_{\pi N}$ is plotted in Fig.~\ref%
{abg-cad}(a).

\begin{figure}[htb]
\centerline{%
\epsfig{file=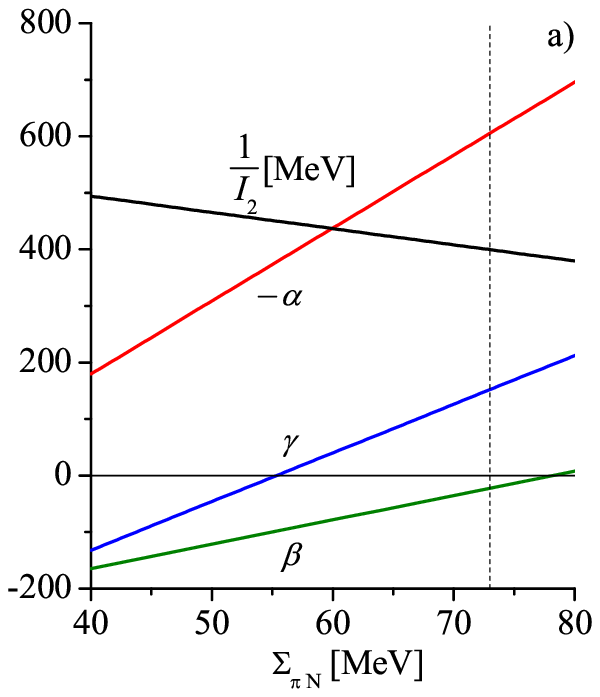,width=6cm} %
\epsfig{file=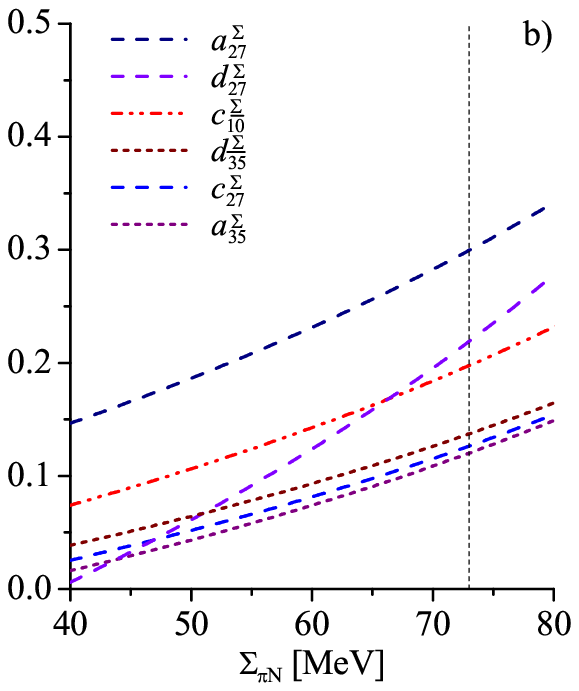,width=6cm}}

\caption{Parameters of the splitting Hamiltonian (%
\ref{Hprim}) and $1/I_{2}$ (a) and (b) mixing coefficients (\ref{mix}) for $%
{\it\Sigma}$ as functions of pion--nucleon sigma term ${\it\Sigma}_{\protect\pi N}$.}
\label{abg-cad}
\end{figure}

Hamiltonian (\ref{Hprim}) introduces mixing between different
representations\break \cite{oper,EKP}:
\vspace{-0.3cm}
\begin{align}
\left| B_{8}\right\rangle & =\left| 8_{1/2},B\right\rangle +c_{\overline{10}%
}^{B}\left| \overline{10}_{1/2},B\right\rangle +c_{27}^{B}\left|
27_{1/2},B\right\rangle \,,  \notag \\
\left| B_{10}\right\rangle & =\left| 10_{3/2},B\right\rangle
+a_{27}^{B}\left| 27_{3/2},B\right\rangle +a_{35}^{B}\left|
35_{3/2},B\right\rangle \,,  \notag \\
\left| B_{\overline{10}}\right\rangle & =\left| \overline{10}%
_{1/2},B\right\rangle +d_{8}^{B}\left| 8_{1/2},B\right\rangle
+d_{27}^{B}\left| 27_{1/2},B\right\rangle +d_{\overline{35}}^{B}\left|
\overline{35}_{1/2},B\right\rangle\,,  \label{admix}
\end{align}%
where $\left| B_{\mathcal{R}}\right\rangle $ denotes the state which reduces
to the SU(3) representation $\mathcal{R}$ in the formal limit $%
m_{s}\rightarrow 0$. The $m_{s}$ dependent (through the linear $m_{s}$
dependence of $\alpha $, $\beta $ and $\gamma $) coefficients in Eq.~(\ref%
{admix}) read:
\begin{eqnarray}
c_{\overline{10}}^{B}& =&c_{\overline{10}}\left[ \kern-0.5em%
\begin{array}{c}
\sqrt{5} \\
0 \\
\sqrt{5} \\
0%
\end{array}%
\kern-0.2em\right] \kern-0.2em,\quad c_{27}^{B}=c_{27}\left[ \kern-0.5em%
\begin{array}{c}
\sqrt{6} \\
3 \\
2 \\
\sqrt{6}%
\end{array}%
\kern-0.2em\right] \kern-0.2em,\;
\nonumber \\
a_{27}^{B}&=&a_{27}\left[ \kern-0.5em%
\begin{array}{c}
\sqrt{15/2} \\
2 \\
\sqrt{3/2} \\
0%
\end{array}%
\kern-0.2em\right] \kern-0.2em,\quad
a_{35}^{B}=a_{35}\left[ \kern-0.5em%
\begin{array}{c}
5/\sqrt{14} \\
2\sqrt{5/7} \\
3\sqrt{5/14} \\
2\sqrt{5/7}%
\end{array}%
\kern-0.2em\right],  \nonumber \\
d_{8}^{B}& =&d_{8}\left[
\begin{array}{c}
0 \\
\sqrt{5} \\
\sqrt{5} \\
0%
\end{array}%
\right] ,\quad d_{27}^{B}=d_{27}\left[
\begin{array}{c}
0 \\
\sqrt{3/10} \\
2/\sqrt{5} \\
\sqrt{3/2}%
\end{array}%
\right] ,\quad d_{\overline{35}}^{B}=d_{\overline{35}}\left[
\begin{array}{c}
1/\sqrt{7} \\
3/(2\sqrt{14)} \\
1/\sqrt{7} \\
\sqrt{5/56}%
\end{array}%
\right]\nonumber\\  \label{mix}
\end{eqnarray}%
in the basis $[N,{\it\Lambda} ,{\it\Sigma} ,{\it\Xi} ]$, $[{\it\Delta} ,{\it\Sigma} ^{\ast },{\it\Xi}
^{\ast },{\it\Omega} ]$ and $\left[ {\it\Theta} ^{+},N_{\overline{10}},{\it\Sigma} _{%
\overline{10}},{\it\Xi} _{\overline{10}}\right] $, respectively, and
\begin{align}
c_{\overline{10}}=-\frac{I_{2}}{15}\left( \alpha +\frac{1}{2}\gamma \right)
,& ~c_{27}=-\frac{I_{2}}{25}\left( \alpha -\frac{1}{6}\gamma \right) ,
\notag \\
~~&  \notag \\
a_{27}=-\frac{I_{2}}{8}\left( \alpha +\frac{5}{6}\gamma \right) ,& ~a_{35}=-%
\frac{I_{2}}{24}\left( \alpha -\frac{1}{2}\gamma \right) ,  \notag \\
~~&  \notag \\
d_{8}=\frac{I_{2}}{15}\left( \alpha +\frac{1}{2}\gamma \right) ,~~& d_{27}=-%
\frac{I_{2}}{8}\left( \alpha -\frac{7}{6}\gamma \right) ,~~~d_{\overline{35}%
}=-\frac{I_{2}}{4}\left( \alpha +\frac{1}{6}\gamma \right) .  \label{mix1}
\end{align}%
In Fig.~\ref{abg-cad}(b) we plot the value of mixing coefficients (\ref{mix})
for the ${\it\Sigma} $ particle\footnote{%
Note that ${\it\Sigma} $ mixes in all the above representations, and therefore is
useful for the sake of illustration.}. We see from Fig.~\ref{abg-cad}(b) that
previously neglected \cite{DPP,Arndt} mixing with $27$ both for $10$ and $%
\overline{10}$ are potentially important (the latter one only for not too
small ${\it\Sigma} _{\pi N}$). Whether the entire correction to the decay widths
due to the admixture of $\mathcal{R}=27$ remains large, depends on the
values of the pertinent transition matrix elements which will be calculated
in the next section.

\section{Decay widths}

\label{decays}

In this section we shall calculate matrix elements which enter into the
formula for the decay width for $B\rightarrow B^{\prime}+\varphi$:%
\begin{equation}
{\it\Gamma}_{B\rightarrow B^{\prime}+\varphi}=\frac{1}{8\pi}\frac{p}{M\,M^{\prime}%
}\overline{\mathcal{M}^{2}}=\frac{1}{8\pi}\frac{p^{3}}{M\,M^{\prime}}%
\overline{\mathcal{A}^{2}}  \label{Gammadef}
\end{equation}
up to linear order in $m_{s}$. The ``bar'' over the amplitude squared denotes
averaging over initial and summing over final spin (and, if explicitly
indicated, over isospin). Anticipating linear momentum dependence of the
decay amplitude $\mathcal{M}$ we have introduced reduced amplitude $\mathcal{%
A}$ which does not depend on the kinematics, \ie on the meson momentum $p$
\begin{equation}
p=\frac{\sqrt{(M^{2}-(M^{\prime}-m_{\varphi})^{2})(M^{2}-(M^{\prime
}-m_{\varphi})^{2})}}{2M}\,.
\end{equation}
For the discussion of the validity of (\ref{Gammadef}) see \cite{EKP}. In
order to match former normalization \cite{DPP} we shall define the decay
amplitude as%
\begin{eqnarray}
\mathcal{M}_{B\rightarrow B^{\prime}+\varphi}&=&\left\langle B^{\prime}\right|
\hat{O}_{\varphi}^{(8)}\left| B\right\rangle \nonumber\\
&=&3\left\langle B^{\prime
}\right| G_{0}D_{\varphi i}-G_{1}d_{ibc}\,D_{\varphi b}^{(8)}\hat{S}_{c}-%
\frac{G_{2}}{\sqrt{3}}D_{\varphi8}^{(8)}\hat{S}_{i}\left| B\right\rangle
\times p_{i}\,,  \label{Op}
\end{eqnarray}
where the sum over repeated indices is assumed: $i=1,2,3$ and $b,c=4,\ldots7$%
.

Coupling constants $G_{0,1,2}$ can be related to the elements of the axial
current operators yielding relations \cite{DPP,oper}:%
\begin{equation}
\frac{9}{5}\frac{F}{D}=\frac{G_{0}+\frac{1}{2}G_{1}+\frac{1}{2}G_{2}}{G_{0}+%
\frac{1}{2}G_{1}-\frac{1}{6}G_{2}}  \label{FD}
\end{equation}
and to
\begin{equation}
G_{2}=\frac{2M_{N}}{3F_{\pi}}{\it\Delta}{\it\Sigma}\,,
\end{equation}
where ${\it\Delta}{\it\Sigma}=0.3\pm0.1$ is the ``spin content of the proton'' and $%
F_{\pi}=93$~MeV. Note that formally constants $G_{0}$ and $G_{1,2}$ are of
different order in $N_{\rm c}$, however, as has been shown in \cite{MPGamma}
additional $N_{\rm c}$ dependence comes from the SU(3) Clebsch--Gordan
coefficients \cite{largereps}.

In the following we shall use%
\begin{equation}
G_{1}=\rho\,G_{0}\,,\quad G_{2}=\varepsilon\,G_{0}\,.  \label{rhoeps}
\end{equation}
Equation (\ref{FD}) introduces relation between $\rho$ and $\varepsilon$:%
\begin{equation}
\varepsilon=\frac{9\frac{F}{D}-5}{9\frac{F}{D}+5}(\rho+2)\,.
\end{equation}
Throughout this paper we fix $F/D=0.59$ following Ref.~\cite{EKP}.
\begin{figure}[htb]
\centerline{%
\epsfig{file=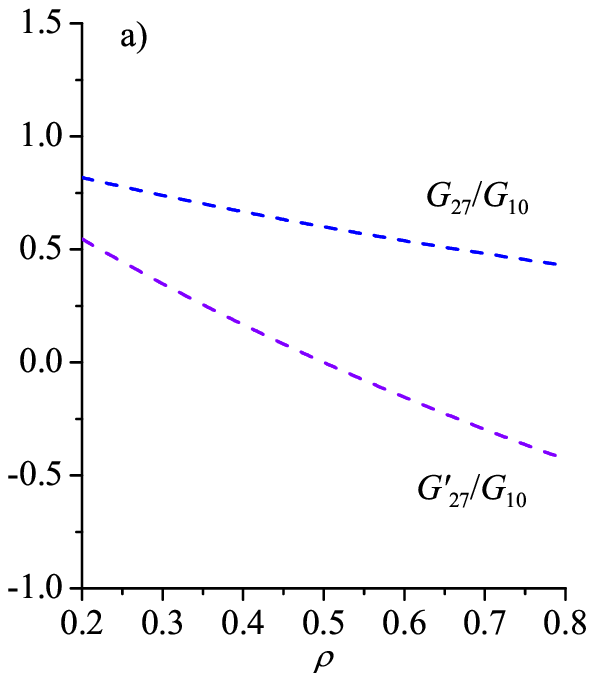,width=5.5cm} %
\epsfig{file=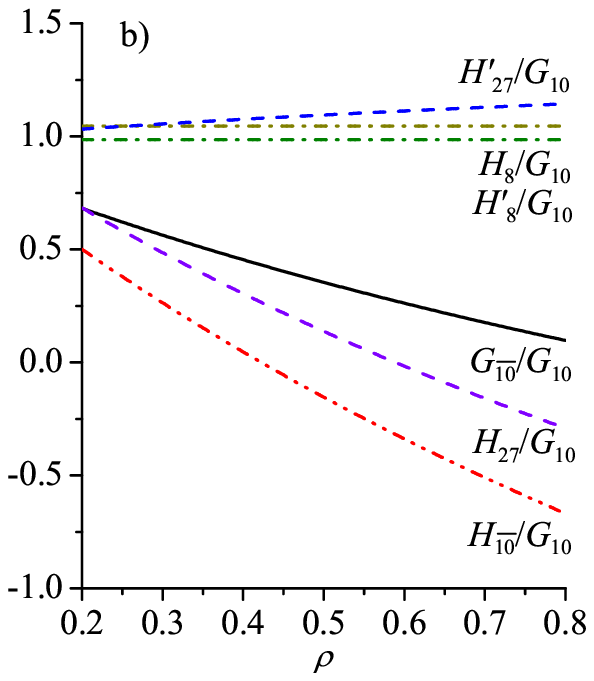,width=5.5cm}}

\vspace{-0.2cm}
\caption{Ratios of the effective couplings (a) (%
\ref{Gs}) and (b) (\ref{HGs}) to $G_{10}$ as functions of the parameter $%
\protect\rho$ defined in Eq.~(\ref{rhoeps}).}
\label{Ratios}

\vspace{-0.2cm}
\end{figure}

\subsection{Decuplet decays}


Now let us consider matrix elements of the decay operator (\ref{Op}) between
states (\ref{admix}). Decuplet can only decay to octet, and we have%
\begin{eqnarray}
\left\langle B_{8}^{\prime}\right| \hat{O}_{\varphi}^{(8)}\left|
B_{10}\right\rangle & =&\left\langle 8_{1/2},B^{\prime}\right| \hat{O}%
_{\varphi}^{(8)}\left| 10_{3/2},B\right\rangle  \notag \\
&& +a_{27}^{B}\left\langle 8_{1/2},B^{\prime}\right| \hat{O}%
_{\varphi}^{(8)}\left| 27_{3/2},B\right\rangle \nonumber\\
&&+c_{27}^{B^{\prime}}\left\langle 27_{1/2},B^{\prime}\right| \hat{O}%
_{\varphi}^{(8)}\left| 10_{3/2},B\right\rangle \,.
\end{eqnarray}
It is convenient to choose $\vec{p}=(0,0,p)$. Then the matrix
elements for $S_{3}=S_{3}^{\prime}=1/2$ read:%
\begin{align}
\left\langle 8_{1/2},B^{\prime}\right| \hat{O}_{\varphi}^{(8)}\left|
10_{3/2},B\right\rangle & =3\frac{2}{\sqrt{15}}G_{10}\left(
\begin{array}{cc}
8 & 8 \\
\varphi & B^{\prime}%
\end{array}
\right| \left.
\begin{array}{c}
10 \\
B%
\end{array}
\right) \times p\,,  \notag \\
\left\langle 8_{1/2},B^{\prime}\right| \hat{O}_{\varphi}^{(8)}\left|
27_{3/2},B\right\rangle & =3\frac{2\sqrt{2}}{9}G_{27}\left(
\begin{array}{cc}
8 & 8 \\
\varphi & B^{\prime}%
\end{array}
\right| \left.
\begin{array}{c}
27 \\
B%
\end{array}
\right) \times p\,,  \notag \\
\left\langle 27_{1/2},B^{\prime}\right| \hat{O}_{\varphi}^{(8)}\left|
10_{3/2},B\right\rangle & =3\frac{1}{\sqrt{15}}G_{27}^{\prime}\left(
\begin{array}{cc}
8 & 27 \\
\varphi & B^{\prime}%
\end{array}
\right| \left.
\begin{array}{c}
10 \\
B%
\end{array}
\right) \times p\,,  \label{dec10}
\end{align}
where%
\begin{equation}
G_{10}=G_{0}+\frac{1}{2}G_{1},\;G_{27}=G_{0}-\frac{1}{2}G_{1},\;G_{27}^{%
\prime}=G_{0}-2G_{1}\,.  \label{Gs}
\end{equation}

We see here that the transition matrix elements are generally different and
depend on representations. For the decuplet decays we find that%
\begin{equation}
G_{10}>G_{27}>G_{27}^{\prime}\sim0\,.
\end{equation}
Hence, we do not expect very large modifications of decuplet decays widths.
Most of the effect will come from $27$ admixture in the initial $10$ state
rather than in the final octet state. However, the admixture coefficient $%
a_{27}$ is relatively large, as can be seen from Fig.~\ref{abg-cad}(b), and
the enhancement factor due to the representation mixing will be of the order
(for $\rho=0.5$ and ${\it\Sigma}_{\pi N}=$ 73~MeV)
\vspace{-0.5cm}
\begin{equation}
R_{B\rightarrow B^{\prime}+\varphi}^{({\rm mix})}\simeq1+2\,a_{27}^{B}\frac{G_{27}%
}{G_{10}}\times C_{B\rightarrow B^{\prime}+\varphi}\sim1+0.4\times
C_{B\rightarrow B^{\prime}+\varphi}\,,  \label{Rmix1}
\end{equation}
where $C_{B\rightarrow B^{\prime}+\varphi}<1$ is the Clebsch--Gordan factor
for a given decay. In the table below we list possible decay modes and the
relevant matrix elements. It should be stressed that for the consistency of
the $m_{s}$ expansion we should not literary square the matrix elements, but
rather --- as in Eq.~(\ref{Rmix1}) --- keep only $G_{10}^{2}$ and the mixed
term $2G_{10}(c\,G_{27}+c^{\prime}\,G_{27}^{\prime})$, neglecting $%
(c\,G_{27}+c^{\prime}\,G_{27}^{\prime})^{2}$.

\begin{equation}
\begin{array}{c|c}
\text{Decay} & \text{Matrix element }\overline{\mathcal{A}^{2}} \\ \hline
~~ & ~~ \\
{\it\Delta}\rightarrow N+\pi & \frac{3}{5}\left[ G_{10}+\frac{5}{3}a_{27}\,G_{27}+%
\frac{1}{3}c_{27}\,G_{27}^{\prime}\right] ^{2} \\
~~ & ~~ \\
{\it\Sigma}_{10}\rightarrow{\it\Lambda}+\pi & \frac{3}{10}\left[ G_{10}+\frac{4}{3}%
a_{27}G_{27}+\frac{2}{3}c_{27}G_{27}^{\prime}\right] ^{2} \\
~~ & ~~ \\
{\it\Sigma}_{10}\rightarrow{\it\Sigma}+\pi & \frac{3}{15}\left[ G_{10}+c_{27}G_{27}^{%
\prime}\right] ^{2} \\
~~ & ~~ \\
{\it\Xi}_{10}\rightarrow{\it\Xi}+\pi & \frac{3}{10}\left[ G_{10}+\frac{7}{9}\sqrt{%
\frac{5}{6}}a_{27}G_{27}+\frac{3}{5}\sqrt{\frac{5}{6}}c_{27}G_{27}^{\prime}%
\right] ^{2}%
\end{array}
\label{tab1}
\end{equation}

\begin{figure}[htb]
\centerline{%
\epsfig{file=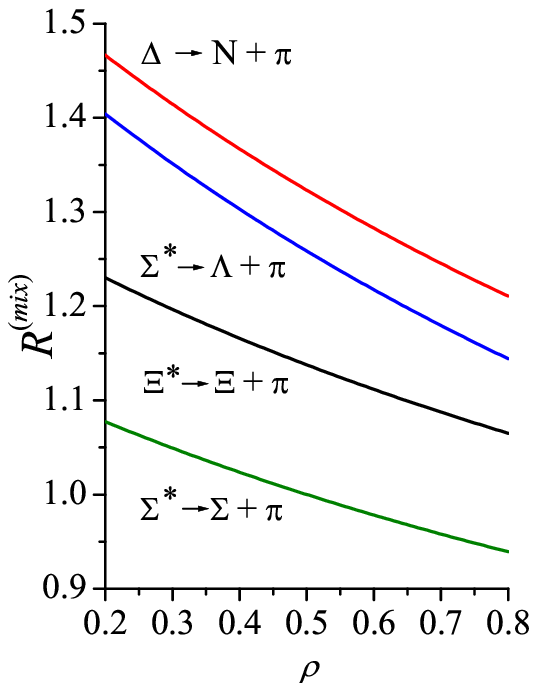,width=5cm} %
\hspace{0.9cm}
\epsfig{file=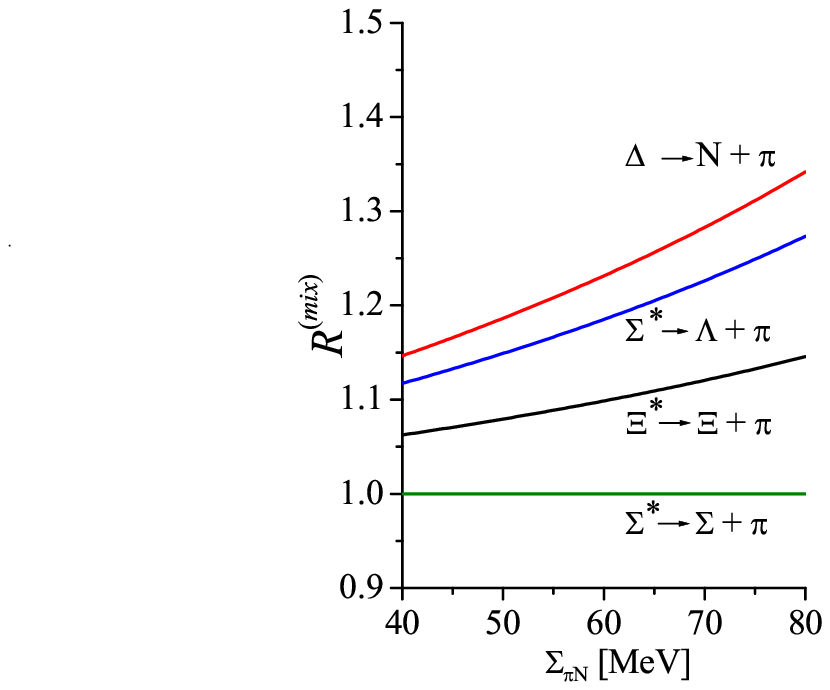,width=5cm}}

\vspace{-0.2cm}
\caption{Enhancement factors $R^{({\rm mix})}$ for the
decuplet decays displayed in Eq.~(\ref{tab1}) as functions of $\protect\rho$
(for ${\it\Sigma}_{\protect\pi N}=73$ MeV) and ${\it\Sigma}_{\protect\pi N}$ (for $%
\protect\rho=0.5$).}
\label{R10}
\end{figure}

In Fig.~\ref{R10} the correction factors
\begin{equation}
R^{({\rm mix})}=1+2\ \frac{\sqrt{\overline{\mathcal{A}^{2}}}-\left. \sqrt{%
\overline{\mathcal{A}^{2}}}\right| _{m_{s}=0}}{\left. \sqrt{\overline{%
\mathcal{A}^{2}}}\right| _{m_{s}=0}}
\end{equation}%
are plotted as functions of $\rho $ and ${\it\Sigma} _{\pi N}$ for the decuplet
decays displayed in Eq.~(\ref{tab1}). We see that they are moderately large
for ${\it\Delta} $ and ${\it\Sigma} ^{\ast }\rightarrow {\it\Lambda} +\pi $. For ${\it\Sigma}
^{\ast }\rightarrow {\it\Sigma} +\pi $ the correction is small since it proceeds
only through the admixture of $27$ in the final state ${\it\Sigma} $, \ie is
given entirely in terms of $G_{27}^{\prime }$\footnote{%
Note that the relevant SU(3) isoscalar factor vanishes for the admixture of
27 in the initial decuplet ${\it\Sigma} ^{\ast }$.}.

\subsection{Antidecuplet decays}

Antidecuplet can directly decay only to octet:%
\begin{eqnarray}
\left\langle B_{8}^{\prime }\right| \hat{O}_{\varphi }^{(8)}\left| B_{%
\overline{10}}\right\rangle & =&\left\langle 8_{1/2},B^{\prime }\right|
\hat{O}_{\varphi }^{(8)}\left| \overline{10}_{1/2},B\right\rangle  \nonumber \\
&& +c_{\overline{10}}^{B^{\prime }}\left\langle \overline{10}_{1/2},B^{\prime
}\right| \hat{O}_{\varphi }^{(8)}\left| \overline{10}_{1/2},B\right\rangle\nonumber \\
&&+c_{27}^{B^{\prime }}\left\langle 27_{1/2},B^{\prime }\right| \hat{O}%
_{\varphi }^{(8)}\left| \overline{10}_{1/2},B\right\rangle  \nonumber \\
&& +d_{8}^{B}\,\left\langle 8_{1/2},B^{\prime }\right| \hat{O}_{\varphi
}^{(8)}\left| 8_{1/2},B\right\rangle \nonumber \\
&&+ d_{27}^{B}\,\left\langle
8_{1/2},B^{\prime }\right| \hat{O}_{\varphi }^{(8)}\left|
27_{1/2},B\right\rangle .  \label{10barmel}
\end{eqnarray}%
In Ref.~\cite{DPP} only the terms proportional to $c_{\overline{10}%
}^{B^{\prime }}$ and $d_{8}^{B}=-c_{\overline{10}}^{B}$ were considered.
Moreover the assumption was made that all transition elements were equal to $%
G_{\overline{10}}$ defined below. Although this is true in the leading order
in the ``explicit'' $N_{\rm c}$ counting\footnote{%
There is also ``implicit'' $N_{\rm c}$ dependence coming from the SU(3)$%
_{\rm flavor}$ Clebsch--Gordan coefficients calculated for an arbitrary $N_{\rm c}$ \cite%
{MPGamma,largereps}.}, the full expressions for these matrix elements are
substantially different from $G_{\overline{10}}$ \cite{EKP}. In Ref.~\cite{Ma}
the admixtures of $27$ were considered with, however, only the leading part
for the transition elements. The comparison with \cite{Ma} can be easily
made by choosing $G_{1}=G_{2}=0,$ $G_{0}=G_{\overline{10}}$ in the equations
below.

In order to evaluate the transition matrix elements entering Eq.~(\ref%
{10barmel}) it is convenient to choose $\vec{p}=(0,0,p)$. Then the
matrix elements for $S_{3}=S_{3}^{\prime}=1/2$ read:%
\begin{eqnarray}
&&\left\langle 8_{1/2},B^{\prime}\right| \hat{O}_{\varphi}^{(8)}\left|
\overline{10}_{1/2},B\right\rangle  =-3\frac{1}{\sqrt{15}}G_{\overline{10}%
}\left(
\begin{array}{cc}
8 & 8 \\
\varphi & B^{\prime}%
\end{array}
\right| \left.
\begin{array}{c}
\overline{10} \\
B%
\end{array}
\right) \times p\,,  \nonumber \\
&&\left\langle \overline{10}_{1/2},B^{\prime}\right| \hat{O}%
_{\varphi}^{(8)}\left| \overline{10}_{1/2},B\right\rangle  =3\frac{1}{2%
\sqrt{6}}H_{\overline{10}}\left(
\begin{array}{cc}
8 & \overline{10} \\
\varphi & B^{\prime}%
\end{array}
\right| \left.
\begin{array}{c}
\overline{10} \\
B%
\end{array}
\right) \times p\,,  \nonumber \\
&&\left\langle 27_{1/2},B^{\prime}\right| \hat{O}_{\varphi}^{(8)}\left|
\overline{10}_{1/2},B\right\rangle  =3\frac{7}{4\sqrt{15}}H_{27}^{\prime
}\left(
\begin{array}{cc}
8 & 27 \\
\varphi & B^{\prime}%
\end{array}
\right| \left.
\begin{array}{c}
\overline{10} \\
B%
\end{array}
\right) \times p\,,  \nonumber 
\\
&&\left\langle 8_{1/2},B^{\prime}\right| \hat{O}_{\varphi}^{(8)}\left|
8_{1/2},B\right\rangle  = \nonumber \\
&&\frac{3}{2}\left[ \frac{H_{8}}{\sqrt{3}}\left(
\begin{array}{cc}
8 & 8 \\
\varphi & B^{\prime}%
\end{array}
\right| \left.
\begin{array}{c}
8_{1} \\
B%
\end{array}
\right) -\sqrt{\frac{3}{5}}H_{8}^{\prime}\left(
\begin{array}{cc}
8 & 8 \\
\varphi & B^{\prime}%
\end{array}
\right| \left.
\begin{array}{c}
8_{2} \\
B%
\end{array}
\right) \right] \times p\,,  \nonumber \\
&&\left\langle 8_{1/2},B^{\prime}\right| \hat{O}_{\varphi}^{(8)}\left|
27_{1/2},B\right\rangle  =-3\frac{1}{9}\sqrt{\frac{2}{5}}H_{27}\left(
\begin{array}{cc}
8 & 8 \\
\varphi & B^{\prime}%
\end{array}
\right| \left.
\begin{array}{c}
27 \\
B%
\end{array}
\right) \times p\,, \label{m_el}
\end{eqnarray}
where we have introduced the following constants\footnote{%
Note that our $H_{27}^{\prime}$ is identical to $H_{\overline{10}}^{\prime}$
from Ref.~\cite{EKP}.} \cite{EKP,Arndt,MPKimmag}:%

\begin{align}
G_{\overline{10}} & =G_{0}-G_{1}-\frac{1}{2}G_{2},\;\;\;\;H_{\overline{10}%
}=G_{0}-\frac{5}{2}G_{1}+\frac{1}{2}G_{2}\,,  \notag \\
H_{27} & =G_{0}-2G_{1}+\frac{3}{2}G_{2},\;\;H_{27}^{\prime}=G_{0}+\frac{11}{%
14}G_{1}+\frac{3}{14}G_{2}\,,  \notag \\
H_{8} & =G_{0}+\frac{1}{2}G_{1}+\frac{1}{2}G_{2},\;\,H_{8}^{\prime}=G_{0}+%
\frac{1}{2}G_{1}-\frac{1}{6}G_{2}\,.  \label{HGs}
\end{align}
In Fig.~\ref{Ratios}(b) we plot the ratios of the transition constants (\ref%
{HGs}) to $G_{10}$. Already here we see the potential source of trouble: the
leading term in (\ref{10barmel}) is governed by $G_{\overline{10}}$ which is
substantially smaller than $G_{10}$. This is the primary source of the
suppression for the antidecuplet decay widths \cite{DPP,EKP}. However, in
view of the smallness of $G_{\overline{10}}$, the coefficients $%
H_{27}^{\prime}$, $H_{8}$ and $H_{8}^{\prime}$, which come from the
admixtures of $27$ in the final octet state and of $8$ in the initial
antidecuplet state, pose a challenge to the validity of the $m_{s}$
expansion, since the relevant mixing coefficients $c_{27}$ and $d_{8}=-c_{%
\overline{10}}$ are not enough suppressed. Therefore we might expect here
relatively large corrections, their magnitude depending on the relative
magnitude of different terms entering the expression for the decay amplitude
$\mathcal{A}$.

\begin{figure}[thb]
\centerline{%
\epsfig{file=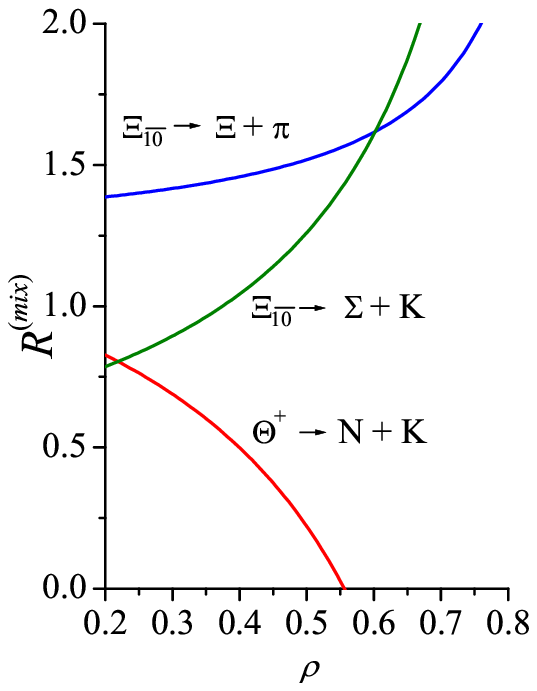,width=5.5cm} %
\epsfig{file=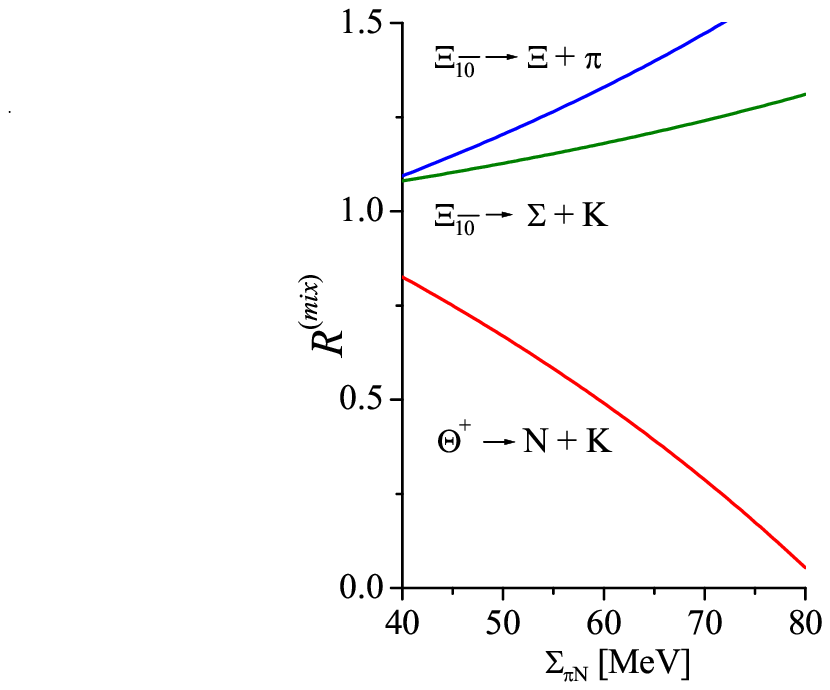,width=5.5cm}}

\caption{Correction coefficients $R^{({\rm mix})}$ for $%
{\it\Theta}^{+}$ and ${\it\Xi}_{\overline{10}}$ decays as functions of $\protect\rho$
(for ${\it\Sigma}_{\protect\pi N}=73$ MeV) and ${\it\Sigma}_{\protect\pi N}$ (for $%
\protect\rho=0.5$).}
\label{RTh}

\vspace{0.3cm}
\end{figure}

Let us first list formulae for the decay widths of the (possibly) observed
states ${\it\Theta} _{\overline{10}}^{+}$ and ${\it\Xi} _{\overline{10}}$:%
\begin{equation}
\begin{array}{c|c}
\text{Decay} & \text{Matrix element }\overline{\mathcal{A}^{2}} \\ \hline
&  \\
{\it\Theta} ^{+}\rightarrow \left\{
\begin{array}{c}
n+K^{+} \\
p+K^{0}%
\end{array}%
\right. & \frac{3}{10}\left[ G_{\overline{10}}+\frac{5}{4}c_{\overline{10}%
}\,H_{\overline{10}}-\frac{7}{4}c_{27}H_{27}^{\prime }\right] ^{2} \\
&  \\
{\it\Xi} _{\overline{10}}\rightarrow {\it\Xi} +\pi & \frac{3}{10}\left[ G_{\overline{10%
}}+\frac{7}{6}c_{27}\,H_{27}^{\prime }+\frac{1}{3}d_{27}H_{27}\right] ^{2}
\\
&  \\
{\it\Xi} _{\overline{10}}\rightarrow {\it\Sigma} +K & \frac{3}{10}\left[ G_{\overline{%
10}}-\frac{5}{4}c_{\overline{10}}\,H_{\overline{10}}+\frac{7}{12}%
c_{27}\,H_{27}^{\prime }-\frac{1}{3}d_{27}\,H_{27}\right] ^{2}%
\end{array}
\label{tabTh}
\end{equation}%
In Eq.~(\ref{tabTh}) for ${\it\Xi} _{\overline{10}}$ we list expressions for the
averaged decay widths (like ${\it\Xi} _{\overline{10}}\rightarrow {\it\Xi} +\pi $)
that, however, for this particular case equal to the width of the specific
decays of ${\it\Xi} _{\overline{10}}^{--}$ (like ${\it\Xi} _{\overline{10}%
}^{--}\rightarrow {\it\Xi} ^{-}+\pi ^{-}$). The relevant correction factors $%
R^{(mix)}$ are plotted in Fig.~\ref{RTh}. We see that ${\it\Theta} ^{+}$ decay
gets strongly suppressed, while ${\it\Xi} _{\overline{10}}$ decays get enhanced %
\cite{EKP}. The strong suppression of ${\it\Gamma} _{{\it\Theta} ^{+}}$ indicates that
for $\rho \gtrsim 0.3$ and/or ${\it\Sigma} _{\pi N}\gtrsim 50$ MeV one cannot
neglect the square of the admixtures which start to dominate. The numerical
analysis of the impact of these and other factors on the physical decay
widths was done in \cite{EKP}. Let us remark here that for ${\it\Theta} ^{+}$
most of the negative correction comes from the 27 admixture, because the $%
\overline{10}$ admixture in the final nucleon is proportional to $H_{%
\overline{10}}$ which is small and changes sign around $\rho \sim 0.4$.
Therefore initial calculation of the $m_{s}$ correction to this decay \cite%
{DPP}, which showed enhancement rather than suppression, indicates how
important are the terms nonleading in $N_{\rm c}$ as well as the corrections due to
the flavor\break 27-plet. The importance of the 27 admixtures is also
visible in the case of ${\it\Xi} _{\overline{10}}\rightarrow {\it\Xi} +\pi $ where all
other admixtures are absent.

Next, let us list results for the nucleon-like states%
\begin{equation}
\arraycolsep1pt
\begin{array}{c|c}
\text{Decay} & \text{Matrix element }\overline{\mathcal{A}^{2}} \\ \hline
&  \\
N_{\overline{10}}\rightarrow\! N+\pi & \frac{3}{20}\left[ G_{\overline{10}}+c_{%
\overline{10}}\left( \frac{5}{4}H_{\overline{10}}-\frac{5}{2}H_{8}-\frac{9}{2%
}H_{8}^{\prime }\right) +c_{27}\frac{49}{12}H_{27}^{\prime }+\frac{1}{15}%
d_{27}H_{27}\right] ^{2} \\
&  \\
N_{\overline{10}}\rightarrow\! N+\eta & \frac{3}{20}\left[ G_{\overline{10}%
}-c_{\overline{10}}\left( \frac{5}{4}H_{\overline{10}}-\frac{5}{2}H_{8}+%
\frac{3}{2}H_{8}^{\prime }\right) -\frac{7}{4}c_{27}H_{27}^{\prime }-\frac{1%
}{5}d_{27}H_{27}\right] ^{2} \\
&  \\
N_{\overline{10}}\rightarrow\! {\it\Lambda} +K & \frac{3}{20}\left[ G_{\overline{10}%
}+c_{\overline{10}}\left( \frac{5}{2}H_{8}+\frac{3}{2}H_{8}^{\prime }\right)
-\frac{7}{2}c_{27}H_{27}^{\prime }+\frac{1}{5}d_{27}H_{27}\right] ^{2} \\
&  \\
N_{\overline{10}}\rightarrow\! {\it\Sigma} +K & \frac{3}{20}\left[ G_{\overline{10}%
}+c_{\overline{10}}\left( \frac{5}{2}H_{\overline{10}}-\frac{5}{2}H_{8}+%
\frac{9}{2}H_{8}^{\prime }\right) -\frac{7}{3}c_{27}H_{27}^{\prime }-\frac{1%
}{15}d_{27}H_{27}\right] ^{2}\\[2mm]%
\end{array}
\label{tabN}
\end{equation}%
which were discussed in some detail in Ref.~\cite{Arndt}. Our formulae agree
with the ones of Ref.~\cite{Arndt} provided we neglect $27$ admixtures and
set $H_{8}=H_{8}^{\prime }$ which in view of Fig.~\ref{Ratios}(b) is
legitimate. At the end of this section we shall discuss the relative
magnitude of the $27$ admixtures with respect to the other ones. Note that
for our set of parameters (\ref{albega}) for ${\it\Sigma} _{\pi N}=73$ MeV $N_{%
\overline{10}}$ (1646~MeV) is below the threshold for ${\it\Sigma} +K$.
\begin{figure}[htb]
\centerline{%
\epsfig{file=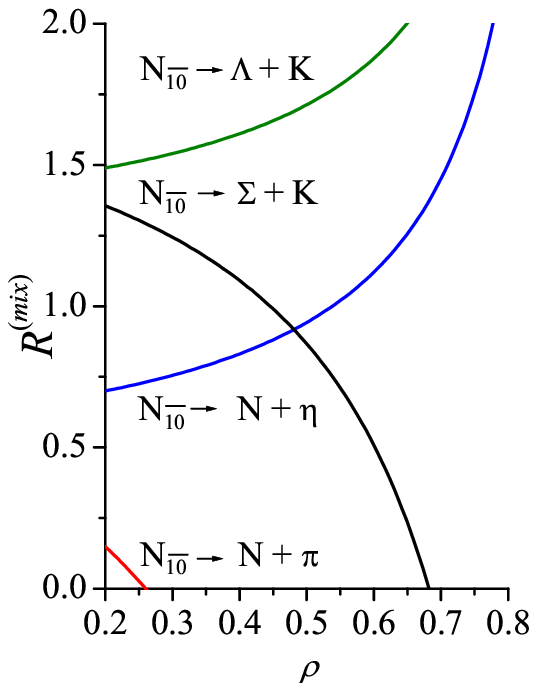,width=5cm} %
\hspace{0.9cm}
\epsfig{file=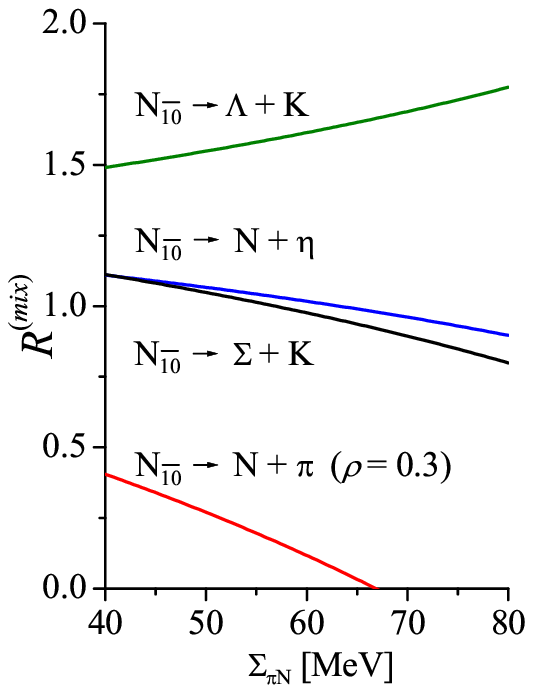,width=5cm}}

\caption{Correction coefficients $R^{({\rm mix})}$ for $N_{%
\overline {10}}$ decays as functions of $\protect\rho$ (for ${\it\Sigma}_{\protect%
\pi N}=73$ MeV) and ${\it\Sigma}_{\protect\pi N}$ (for $\protect\rho=0.5$ and $%
\protect\rho=0.3$).}
\label{RN}

\vspace{-0.3cm}
\end{figure}

\begin{figure}[htb]
\centerline{%
\epsfig{file=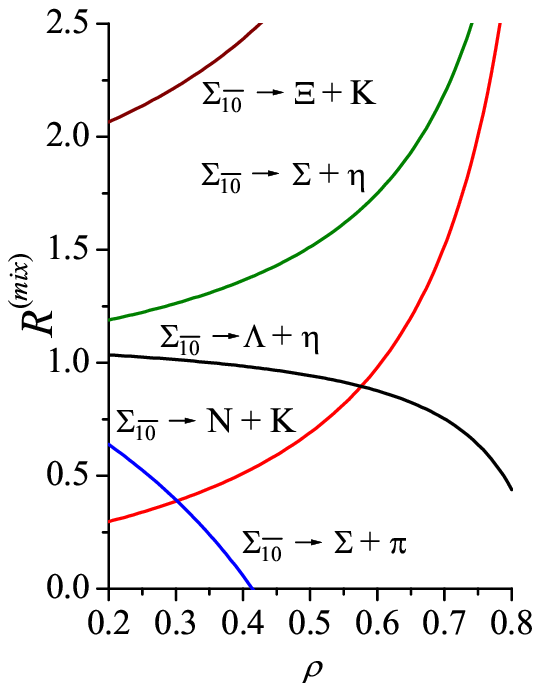,width=6cm} %
\epsfig{file=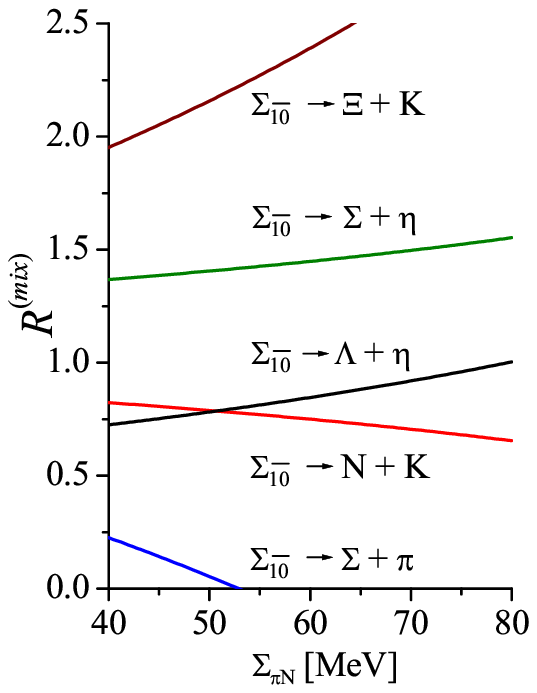,width=6cm}}

\caption{Correction coefficients $R^{({\rm mix})}$ for $%
{\it\Sigma} _{\overline{10}}$ decays as functions of $\protect\rho $ (for ${\it\Sigma}
_{\protect\pi N}=$ 73~MeV) and ${\it\Sigma} _{\protect\pi N}$ (for $\protect\rho %
=0.5$).}
\label{RS}
\end{figure}

For ${\it\Sigma} _{\overline{10}}$ decays we get:%
\begin{equation}
\arraycolsep2pt
\begin{array}{c|c}
\text{Decay} & \text{Matrix element }\overline{\mathcal{A}^{2}} \\ \hline
&  \\
{\it\Sigma} _{\overline{10}}\rightarrow N+K & \frac{1}{10}\left[ G_{\overline{10}%
}-c_{\overline{10}}\left( \frac{5}{2}H_{\overline{10}}-\frac{5}{2}H_{8}+%
\frac{9}{2}H_{8}^{\prime }\right) +\frac{7}{6}c_{27}H_{27}^{\prime }-\frac{4%
}{15}H_{27}\right] ^{2} \\
&  \\
{\it\Sigma} _{\overline{10}}\rightarrow {\it\Sigma} +\pi & \frac{1}{10}\left[ G_{%
\overline{10}}+c_{\overline{10}}\left( \frac{5}{2}H_{\overline{10}%
}-5H_{8}\right) +\frac{7}{2}c_{27}H_{27}^{\prime }\right] ^{2} \\
&  \\
{\it\Sigma} _{\overline{10}}\rightarrow {\it\Sigma} +\eta & \frac{3}{20}\left[ G_{%
\overline{10}}+3c_{\overline{10}}H_{8}^{\prime }-\frac{7}{3}%
c_{27}H_{27}^{\prime }-\frac{4}{15}d_{27}H_{27}\right] ^{2} \\
&  \\
{\it\Sigma} _{\overline{10}}\rightarrow {\it\Lambda} +\eta & \frac{3}{20}\left[ G_{%
\overline{10}}-3c_{\overline{10}}H_{8}^{\prime }+\frac{7}{2}%
c_{27}H_{27}^{\prime }+\frac{4}{15}d_{27}H_{27}\right] ^{2} \\
&  \\
{\it\Sigma} _{\overline{10}}\rightarrow {\it\Xi} +K & \frac{3}{30}\left[ G_{\overline{%
10}}+c_{\overline{10}}\left( \frac{5}{2}H_{8}+\frac{9}{2}H_{8}^{\prime
}\right) -\frac{14}{3}c_{27}H_{27}^{\prime }+\frac{4}{15}d_{27}H_{27}\right]
^{2}\\[2mm]%
\end{array}
\label{tabSig}
\end{equation}%
and the pertinent correction factors are plotted in Fig.~\ref{RS}. Note that
for our set of parameters (\ref{albega}) for ${\it\Sigma} _{\pi N}=73$ MeV $%
{\it\Sigma} _{\overline{10}}$ (1754 MeV) is below the threshold for ${\it\Xi} +K$.

Finally let us discuss the importance of the 27 admixtures both in the
initial and final state. For decuplet decays this is the only possible $%
m_{s} $ correction, however, the final state admixture is proportional to $%
G_{27}^{\prime}$ which small. The magnitude of these corrections can be read
off from Fig.~\ref{R10}. For antidecuplet decays only ${\it\Xi}_{\overline{10}%
}\rightarrow{\it\Xi}+\pi$ is entirely governed by $\mathcal{R}=27$ $m_{s}$
corrections, while for the other decays there is a subtle interplay of 27
and $\overline{10}+8$, however, in view of Fig.~\ref{Ratios}, the final
state 27 admixture is by far more important that the one in the initial
state. In Eq.~(\ref{tab27}) below we present the ratios of the $\overline{10}$
and $8$ admixtures to $27$ for $\rho=0.5$ and ${\it\Sigma}_{\pi N}=73$ MeV:%
\begin{equation}
\begin{array}{c|c}
\text{Decay} & \text{Ratio (}\overline{10}+8\text{) to }27 \\ \hline
{\it\Theta}^{+}\rightarrow N+K\w & \;\;0.14 \\
{\it\Xi}_{\overline{10}}\rightarrow{\it\Sigma}+K & \;\;0.58 \\
N_{\overline{10}}\rightarrow N+\pi & -2.17 \\
N_{\overline{10}}\rightarrow N+\eta & -0.89 \\
N_{\overline{10}}\rightarrow{\it\Lambda}+K & -1.48 \\
N_{\overline{10}}\rightarrow{\it\Sigma}+K & -0.83 \\
{\it\Sigma}_{\overline{10}}\rightarrow N+K & -1.70 \\
{\it\Sigma}_{\overline{10}}\rightarrow{\it\Sigma}+\pi & -1.98 \\
{\it\Sigma}_{\overline{10}}\rightarrow{\it\Sigma}+\eta & -1.48 \\
{\it\Sigma}_{\overline{10}}\rightarrow{\it\Lambda}+\eta & -1.01 \\
{\it\Sigma}_{\overline{10}}\rightarrow{\it\Xi}+K & -1.91%
\end{array}
\label{tab27}
\end{equation}

\newpage\noindent
An interesting pattern emerges: ${\it\Theta}^{+}$ and ${\it\Xi}_{\overline{10}}$
decays are dominated by $\mathcal{R}=27$ $m_{s}$ corrections,
whereas for $N_{\overline{10}}$ and for ${\it\Sigma}_{\overline{10}}$ 
decays $\overline{10}+8$ and $27$ pieces are comparable, 
with $27$ admixture beeing at most 2 times smaller and of the opposite sign.

\section{Summary}

\label{summ}
\vspace{0.3cm}
The representation mixing introduced by the symmetry breaking Hamiltonian (%
\ref{Hprim}) depends on the value of the ${\it\Sigma} _{\pi N}$ term. For ${\it\Sigma}
_{\pi N}=73$ MeV, the value suggested by recent experimental analysis \cite%
{sigma}, as well as by the chiral soliton model fits \cite{EKP,Schw}, this
mixing becomes large and introduces large ${\mathcal{O}}(m_{s})$ corrections
to the decay widths. In the case of the decuplet decays large means at most $%
35$\%, however, in the antidecuplet case these corrections may reach even a
factor 3 or more. This is because the leading term, given by the transition
matrix element $G_{\overline{10}}$ of Eq.~(\ref{HGs}), is small for the
typical values of the coupling constants $G_{0,1,2}$ entering the decay
operator (\ref{Gammadef}). This poses a challenge to the SU(3) perturbation
expansion in $m_{s}$. As far as two exotic states recently observed
experimentally \cite{exp,Xi} are concerned, we find strong suppression of
the ${\it\Theta} ^{+}$ decay width and the enhancement for ${\it\Xi} _{\overline{10}}$%
. Before, however, a definitive conclusion may be drawn, the ${\mathcal{O}}%
(m_{s})$ corrections to the decay operator (\ref{Gammadef}) must be examined.

A key issue is whether, knowing the decay width of ${\it\Delta} $, one can
predict the decay width of ${\it\Theta} ^{+}$. Other decays of the antidecuplet
are in fact related to ${\it\Theta} ^{+}$ by phase space factors, for which we
have to know the masses of the decaying particles, and the SU(3)
Clebsch--Gordan coefficients and the $\mathcal{O}(m_{s})$ correction factors
calculated in this paper. In order to estimate ${\it\Gamma} _{{\it\Theta}
^{+}\rightarrow n+K^{+}}$ let us observe that%
\begin{equation}
\left. \frac{{\it\Gamma} _{{\it\Theta} ^{+}\rightarrow n+K^{+}}}{{\it\Gamma} _{{\it\Delta} }}%
\right| _{m_{s}=0}=\frac{1}{2}\frac{M_{{\it\Delta} }}{M_{{\it\Theta} ^{+}}}\left(
\frac{p_{K}}{p_{\pi }}\right) ^{3}\left( \frac{G_{\overline{10}}}{G_{10}}%
\right) ^{2}=0.67\left( \frac{G_{\overline{10}}}{G_{10}}\right) ^{2}\,,
\end{equation}%
where $G_{\overline{10}}/G_{10}$ is plotted in Fig. \ref{Ratios}(b). For $%
\rho \sim 0.5$ we get ${\it\Gamma} _{{\it\Theta} ^{+}\rightarrow n+K^{+}}/{\it\Gamma}
_{{\it\Delta} }\sim 0.1$, a fairly large suppression\footnote{%
Note that we discuss here one of the two possible decay modes of ${\it\Theta}
^{+} $, rather than the total width, which would make this ratio $0.2$.}.
Next, the $\mathcal{O}(m_{s})$ corrections increase ${\it\Delta} $ width by a
factor of $1.3$ and suppress ${\it\Gamma} _{{\it\Theta} ^{+}}$ by a factor of $0.2$,
so that the relative suppression coming from the SU(3) breaking is $0.15$.
As said above this factor must be considered with care, however, the
suppression mechanism is here evident and one may safely conclude that%
\begin{equation}
\frac{{\it\Gamma} _{{\it\Theta} ^{+}}}{{\it\Gamma} _{{\it\Delta} }}\ll 0.1
\end{equation}%
indicating that the total width ${\it\Gamma} _{{\it\Theta} ^{+}}$ is of the order
of a few MeV. Further suppression factors were discussed in the
original paper of Diakonov, Petrov and Polyakov \cite{DPP} and more recently
in \cite{EKP}.

In order to estimate the widths of the other members of antidecuplet we
first calculate the ratios of a given decay width to the one of ${\it\Theta} ^{+}$
and then the $\mathcal{O}(m_{s})$ correction factors (for $\rho =0.5$ and $%
{\it\Sigma} _{\pi N}=73$ MeV). The results are listed in (\ref{taba10}) below:%
\begin{equation}
\begin{array}{c|ccc}
\text{Decay }X &
\begin{array}{c}
R_{{\it\Gamma} }= \\
{\it\Gamma} /{\it\Gamma} _{{\it\Theta} ^{+}\rightarrow n+K^{+}}%
\end{array}
&
\begin{array}{c}
R_{\rm min}= \\
R^{\rm (mix)}%
\end{array}
&
\begin{array}{c}
R_{\rm max}= \\
R^{({\rm mix})}/R_{{\it\Theta} ^{+}\rightarrow n+K^{+}}^{({\rm mix})}\\[3mm]%
\end{array}
\\ 
\hline
{\it\Theta} ^{+}\rightarrow n+K^{+} & 1 & 0.20 &\w 1 \\
{\it\Xi} _{\overline{10}}\rightarrow {\it\Xi} +\pi  & 2.67 & 1.54 & 7.7 \\
{\it\Xi} _{\overline{10}}\rightarrow {\it\Sigma} +K & 1.57 & 1.27 & 6.4 \\
N_{\overline{10}}\rightarrow N+\pi  & 3.82 & ? & ? \\
N_{\overline{10}}\rightarrow N+\eta  & 0.97 & 0.92 & 4.6 \\
N_{\overline{10}}\rightarrow {\it\Lambda} +K & 0.08 & 1.66 & 8.3 \\
{\it\Sigma} _{\overline{10}}\rightarrow N+K & 1.74 & 0.70 & 3.5 \\
{\it\Sigma} _{\overline{10}}\rightarrow {\it\Sigma} +\pi  & 3.47 & ? & ? \\
{\it\Sigma} _{\overline{10}}\rightarrow {\it\Sigma} +\eta  & 2.10 & 1.48 & 7.4 \\
{\it\Sigma} _{\overline{10}}\rightarrow {\it\Lambda} +\eta  & 0.34 & 0.99 & 4.9%
\end{array}
\label{taba10}
\end{equation}%

\vspace{0.3cm}\noindent
The question marks in Eq.~(\ref{taba10}) indicate that the $R^{({\rm mix})}$
factors are negative so that our results are not reliable. Since also the
large suppression factor for ${\it\Theta} ^{+}$ is not fully reliable, one could
argue that the real correction factor lies somewhere between $R_{{\rm min}}$ and $%
R_{{\rm max}}$ defined in (\ref{taba10}). Our estimates for the decay width in the
antidecuplet take therefore the following form%
\begin{equation}
{\it\Gamma} _{X}={\it\Gamma} _{{\it\Theta} ^{+}\rightarrow n+K^{+}}\times R_{{\it\Gamma} }\times
\left\{
\begin{array}{c}
R_{\rm max} \\
\\
R_{\rm min}%
\end{array}%
\right.
\end{equation}%
and for the entries with question marks one can only conclude that%
\begin{equation}
{\it\Gamma} _{X}<{\it\Gamma} _{{\it\Theta} ^{+}\rightarrow n+K^{+}}
\times R_{{\it\Gamma} }\,.
\end{equation}%
Hence for $2$ MeV total decay width of ${\it\Theta} ^{+}$ we get for example
${\it\Gamma} _{{\it\Xi} _{\overline{10}}\,\rightarrow {\it\Xi} +\pi }\sim 4\div
20.6$ MeV and ${\it\Gamma}_{{\it\Xi} _{\overline{10}}\rightarrow {\it\Sigma} +K}
\sim 2 \div 10$ MeV. This is more than the value quoted in
Ref.~\cite{Arndt}, however, these authors did not take into
account the 27 admixture which is quite important in this case.
Also for $N_{\overline{10}}\rightarrow N+\pi$ we find much
stronger suppression than \cite{Arndt} due to the fact that for
our set of parameters $c_{\overline{10}}$ is larger and
$H_{\overline{10}}<0$ rather than small and positive.

\newpage

The author would like to thank John Ellis and Marek Karliner for discussions
which stimulated this research. Helpful e-mail exchanges with Dmitri
Diakonov and Maxim Polyakov are acknowledged. The present work is supported
by the Polish State Committee for Scientific Research (KBN) under grant 2 P03B 043
24. This manuscript has been authored under Contract No. DE-AC02-98CH10886
with the U.S. Department of Energy.


\begin{thebibliography}{99}
\bibitem{exp} T.~Nakano \textit{et al.} (LEPS Collaboration), {\it Phys.\ Rev.\
Lett.}\ \textbf{91}, 012002 (2003);
V.V.~Barmin \textit{et al.} (DIANA Collaboration), {\it Phys.\ Atom.\ Nucl.}\
\textbf{66}, 1715 (2003) [{\it Yad.\ Fiz.}\ \textbf{66}, 1763 (2003)];
S.~Stepanyan \textit{et al.} (CLAS Collaboration), 
{\it Phys. Rev. Lett.} {\bf 91}, 252001 (2003 );
%
J.~Barth \textit{et al.} (SAPHIR Collaboration), {\tt hep-ex/0307083};
%
V.~Kubarovsky, S.~Stepanyan and CLAS Collaboration, 
{\it AIP Conf. Proc.} {\bf 698}, 543 (2004);
%
A.E.~Asratyan, A.G.~Dolgolenko, M.A.~Kubantsev, {\tt hep-ex/0309042};
%
V.~Kubarovsky \textit{et al.} (CLAS Collaboration),
{\tt hep-ex/0311046}; %
A.~Airapetian \textit{et al.} (HERMES Collaboration), 
{\it Phys. Lett.} {\bf B585}, 213 (2004);
S.~Chekanov (ZEUS Collaboration), {\tt
http://www.desy.de/f/seminar/Chekanov.pdf}; R.~Togoo \textit{et al.}, {\it Proc.
Mongolian Acad. Sci.} \textbf{4}, 2 (2003);
A.~Aleev \textit{et al.} (SVD Collaboration), {\tt hep-ex/0401024.}

\bibitem{Xi} C.~Alt \textit{et al.} (NA49 Collaboration),
{\it Phys. Rev. Lett.} {\bf 92}, 042003 (2004). 

\bibitem{antidec} P.O.~Mazur, M.A.~Nowak, M.~Prasza{\l}owicz, {\it Phys.\
Lett.}\  \textbf{B147}, 137 (1984); %
A.V.~Manohar,
{\it Nucl.\ Phys.}\  \textbf{B248}, 19 (1984); %
M.~Chemtob, {\it Nucl.\ Phys.}\  \textbf{B256}, 600 (1985);
S.~Jain, S.R.~Wadia, %
{\it Nucl.\ Phys.}\  \textbf{B258}, 713 (1985); %
M.P.~Mattis, M.~Karliner, {\it Phys.\ Rev.}\  \textbf{D31}, 2833 (1985);
M.~Karliner, M.P.~Mattis, {\it Phys.\ Rev.}\  \textbf{D34}, 1991 (1986).

\bibitem{Mogil} M.~Prasza{\l}owicz, \textit{Proc. of the Workshop on
Skyrmions and Anomalies}, Kraków, 1987, eds. M~Je{\.z}abek and M.~Prasza{\l}%
owicz, World Scientific, Singapore 1987, p.~531; M.~Prasza{\l}owicz, {\it Phys.\
Lett.}\  \textbf{B575}, 234 (2003).

\bibitem{DPP} D.~Diakonov, V.~Petrov, M.V.~Polyakov, {\it Z.\ Phys.}\ 
\textbf{A359}, 305 (1997).

\bibitem{Weigel} H.~Weigel, {\it Eur.\ Phys.\ J.}\  \textbf{A2}, 391
(1998);
{\it  AIP Conf. Proc.} {\bf 549}, 271 (2002). %

\bibitem{sigma} R. Koch, {\it Z. Phys.} \textbf{C15}, 161 (1982); W.~Wiedner \emph{%
et al}. {\it Phys. Rev. Lett.} \textbf{58}, 648 (1987); J.~Gasser, H.~Leutwyler,
M.P.~Locher, M.E.~Sainio, {\it Phys. Lett.} \textbf{B213}, 85 (1988);
M.M.~Pavan, I.I.~Strakovsky, R.L.~Workman, R.A.~Arndt, {\it PiN Newslett.}\
\textbf{16}, 110 (2002).
For other recent estimates, see: T.~Inoue, V.E.~Lyubovitskij,
T.~Gutsche,
 A.~Faessler, {\tt hep-ph/0311275} and references therein.

\bibitem{other27} H.~Walliser, V.B.~Kopeliovich, {\it J.\ Exp.\ Theor.\
Phys.}\ \textbf{97}, 433 (2003) [Zh.\ Eksp.\ Teor.\ Fiz.\ \textbf{124}, (2003)
483]; %
D.~Borisyuk, M.~Faber, A.~Kobushkin, {\tt hep-ph/0307370},
{\tt hep-ph/0312213.} %

\bibitem{Ma} B.~Wu, B.Q.~Ma,
{\tt hep-ph/0312041}, {\tt hep-ph/0312326.}
arXiv: %

\bibitem{EKP} J.~Ellis, M.~Karliner, M.~Prasza{\l}owicz,
{\tt hep-ph/0401127.} %

\bibitem{Arndt} R.A.~Arndt, Y.I.~Azimov, M.V.~Polyakov, I.I.~Strakovsky,
 R.L.~Workman,
{\tt nucl-th/0312126.} %

\bibitem{MPGamma} M.~Prasza{\l}owicz,
 {\it Phys. Lett.} \textbf{B583}, 96 (2004).

\bibitem{oper} H.C.~Kim, M.~Prasza{\l}owicz, K.~Goeke,
{\it Phys.\ Rev.}\ \textbf{D61}, 114006 (2000);
H.C.~Kim, M.V.~Polyakov, M.~Prasza{\l}owicz, K.~Goeke,
{\it Phys.\ Rev.} \textbf{D57}, 299 (1998);
H.C.~Kim, M.~Prasza{\l}owicz, K.~Goeke,
{\it Phys.\ Rev.}\ \textbf{D57}, 2859 (1998).

\bibitem{Blotzsu3} A.~Blotz, D.~Diakonov, K.~Goeke, N.W.~Park,
V.~Petrov, P.V.~Pobylitsa,
{\it Nucl.\ Phys.}\ \textbf{A555}, 765 (1993). %

\bibitem{Leutwyler} H.~Leutwyler, {\it Nucl.\ Phys.\ Proc.\ Suppl.}\
\textbf{94}, 108
(2001). %

\bibitem{Schw} P.~Schweitzer,
{\tt hep-ph/0312376.} %

\bibitem{largereps} G. Karl, J. Patera, S. Perantonis, {\it Phys. Lett.} \textbf{%
172B}, 49 (1986); J.~Bijnens, H.~Sonoda, M.~Wise, {\it Can. J. Phys.}
\textbf{64}, 1
(1986); Z.~Duli{\'n}ski, M.~Prasza{\l}owicz, {\it Acta Phys. Pol. B} \textbf{18%
}, 1157 (1988); Z.~Duli{\'n}ski, {\it Acta. Phys. Pol. B} \textbf{19}, 891 (1988).

\bibitem{MPKimmag} H.C.~Kim, M.~Prasza{\l}owicz, {\tt hep-ph/0308242},
to be published in {\it Phys. Lett.}~\textbf{B}.
\end{thebibliography}
\end{document}